# Chiral Dirac-like fermion in spin-orbit-free antiferromagnetic semimetals

Pengfei Liu,[1,2] Ao Zhang,[1] Jingzhi Han,[2] and Qihang Liu[1,3,4,*]

[1]Shenzhen Institute for Quantum Science and Engineering and Department of Physics, Southern University of Science and Technology, Shenzhen 518055, China

[2]School of Physics, Peking University, Beijing 100871, China

[3]Guangdong Provincial Key Laboratory of Computational Science and Material Design, Southern University of Science and Technology, Shenzhen 518055, China

[4]Shenzhen Key Laboratory of Advanced Quantum Functional Materials and Devices, Southern University of Science and Technology, Shenzhen 518055, China

*Correspondence: liuqh@sustech.edu.cn

Dirac semimetal is a phase of matter, whose elementary excitation is described by the relativistic Dirac equation. In the limit of zero mass, its parity-time symmetry enforces the Dirac fermion in the momentum space, which is composed of two Weyl fermions with opposite chirality, to be non-chiral. Inspired by the flavor symmetry in particle physics, we theoretically propose a massless Dirac-like equation yet linking two Weyl fields with the identical chirality by assuming $SU(2)$ isospin symmetry, independent of the space-time rotation exchanging the two fields. Dramatically, such symmetry is hidden in certain solid-state spin-1/2 systems with negligible spin-orbit coupling, where the spin degree of freedom is decoupled with the lattice. Therefore, the existence of the corresponding quasiparticle, dubbed as flavor Weyl fermion, cannot be explained by the conventional (magnetic) space group framework. The four-fold degenerate flavor Weyl fermion manifests linear dispersion and a Chern number of $\pm 2$, leading to a robust network of topologically protected Fermi arcs throughout the Brillouin zone. For material realization, we show that the transition-metal chalcogenide $CoNb_3S_6$ with experimentally confirmed collinear antiferromagnetic order is ideal for flavor Weyl semimetal under the approximation of vanishing spin-orbit coupling. Our work reveals a counterpart of the flavor symmetry in magnetic electronic systems, leading to further possibilities of emergent phenomena in quantum materials.

## INTRODUCTION

The Dirac equation combines the two cornerstones of modern physics—quantum mechanics and relativity. It is the first step towards the quantum field theory that gives birth to the standard model of particle physics. It complies with the Dirac quantum fields of spin-½ particles, furnishing particular irreducible representations (irreps) of the Lorentz group. There are several manifestations of the Dirac equation in condensed matter systems, such as graphene,[1] topological insulators,[2-4] Dirac semimetals (DSMs),[5-7] Weyl semimetals,[8,9] and *d*-wave high-temperature superconductors. The low-energy electronic structure of the Dirac points in a three-dimensional (3D) DSM, i.e., four-fold degenerate crossing points formed by doubly degenerate linear bands, is well described by the massless Dirac equation in the (3+1)D space-time.[6,10] Investigations on 3D DSMs have been largely confined to the field of nonmagnetic materials where inversion symmetry *P* and time-reversal symmetry *T* coexist,[6,11-14] ensuring the doubly degenerate bands constituting the Dirac point. Later, candidates for DSM have been extended to include magnetic materials with broken *T* but preserved *PT* symmetry.[15-17] Recent progresses comprehensively conduct topological classifications of magnetic materials to identify topological nontrivial insulators and semimetals by employing the full magnetic space groups,[18-21] which are also utilized for the construction of $k \cdot p$ models in order to classify emergent quasiparticle excitations in magnetic materials.[22,23]

A four-component Dirac field can be decomposed into two two-component Weyl fields with opposite chirality in the limit of zero mass, implying that the chirality of a massless Dirac fermion must be zero because the *PT*-symmetry forces the two branches of each doubly degenerate band to have opposite Berry curvatures (Figure 1A). Hence, the Fermi arc surface states connecting two Dirac points in a DSM are generally not topologically protected, unlike Weyl points (Figure 1B). These properties establish the current textbook *Gestalt* underlying our understanding of Dirac physics.

Here, we propose the theory and material realization of a new semimetal phase having Dirac-like four-fold degenerate points formed by doubly degenerate bands, yet a nonzero chirality, dubbed as flavor Weyl semimetal (WSM, Figure 1C**,**D). This is achieved by a massless four-component field in vacuum furnishing chiral and four-dimensional (4D) irreps, connecting two Weyl fields via a type of $SU(2)$ flavor symmetry—analogous to the isospin symmetry relating a proton and a neutron. Remarkably, such $SU(2)$ isospin symmetry can be obtained by spin space group—a type of expanded symmetry group compared with the traditional magnetic space group—existing in magnetic materials with negligible spin-orbit coupling (SOC).[24,25] Such groups were originally applied to describe the symmetry of magnons in Heisenberg Hamiltonian, while drawing recent attention for the application in discovering new topological invariants and magnetic topological phases.[26-28] We show that the transition-metal chalcogenide $CoNb_3S_6$ with a chiral crystal structure and collinear antiferromagnetic (AFM) order is an ideal candidate for such flavor WSM. The resulting four-fold degenerate quasiparticles have Chern numbers $C = \pm 2$, manifesting a robust network of topologically protected Fermi arcs throughout the surface Brillouin zone. Furthermore, the modified band property and topology by the effects of SOC indicate that the flavor WSM phase serves as a good starting point to understand the topological nature of $CoNb_3S_6$.

**RESULTS**

**Theory of Dirac-like fermion with chirality**

In high energy physics, a Dirac field is a four-component field with field operators furnishing a 4D irreducible representation of Lorentz group. Such a field is nonchiral because $P$ connects two 2D representations of proper orthrochronous Lorentz group supporting two Weyl fields with opposite chirality. To construct a chiral Dirac-like 4-component field, one requires additional internal symmetries (i.e., commute with space-time operations[29]) connecting two Weyl fields with identical chirality. A famous type of internal symmetry in particle physics is the isospin symmetry pairing a proton and a neutron forming an $SU(2)$ doublet.[29,30] In analogy, we can choose a condensed-matter counterpart of such $SU(2)$ flavor symmetry to construct a massless four-component field with chirality, named flavor Weyl field (see Supplementary Note S1).

In condensed matter solids with elementary excitations (quasiparticles), although spin is an internal degree of freedom of an electron, its rotational operations are completely locked to the rotations of the lattice owing to the relativistic SOC effect. However, the symmetry description of compounds composed of light elements with negligible SOC requires decoupled spin and lattice operations, forming symmetry groups called spin groups.[24-26] We next show that the combination of translation and spin rotation in certain magnetic compounds with long-range magnetic order leads to a hidden $SU(2)$ symmetry group, supporting the emergence of flavor Weyl fermions.

We considered a collinear AFM system belonging to the type-IV Shubnikov space group, as schematically shown in Figure 2A. We used a four-band model with two orbitals separately located at sublattices $A$ and $B$ and the Neel vector along the z-axis to describe such a system. Three elements in the spin space group were considered, including a two-fold spin rotation perpendicular to the magnetic moment followed by a fractional translation symmetry, $u_x^{1/2} = \{U_x(\pi)||E|\tau_{1/2}\}$, $u_y^{1/2}$ and a spin rotation operation along magnetic moments with an infinitesimal rotation angle, $\{U_z(\theta)||E|0\}$, where $U_n(\theta)$, represents pure spin rotation $\theta$ along $\boldsymbol{n}$ axis and $\boldsymbol{\tau_{1/2}}$ denotes the half translation along the a certain axis (see Supplementary Note S2). We can write the general form of the single-electron Hamiltonian as $H(\boldsymbol{k}) = \Sigma_{i,j=0,x,y,z} f_{ij}(\boldsymbol{k})\,\tau_j \otimes \sigma_i$, where $\sigma$ and $\tau$ operate on spin and site DOF, respectively, and $f_{ij}(\boldsymbol{k})$ represents real functions of $\boldsymbol{k}$. After that, the elements can be represented as $u_x^{1/2} = -ie^{-i\boldsymbol{k}\cdot\boldsymbol{\tau}_{1/2}}\tau_x \otimes \sigma_x$, $u_y^{1/2} = -ie^{-i\boldsymbol{k}\cdot\boldsymbol{\tau}_{1/2}}\tau_x \otimes \sigma_y$ and $\{U_z(\theta)||E|0\} = \tau_0 \otimes e^{-i\theta\sigma_z/2}$. We applied these symmetry constraints to the Hamiltonian $H(\boldsymbol{k})$ and obtained the following equation:

$$H(\boldsymbol{k}) = d_0(\boldsymbol{k})\alpha^0 + \Sigma_{i=1,2,3}\, d_i(\boldsymbol{k})\alpha^i, \qquad (2)$$

where $d_0(\boldsymbol{k})$ and $d_i(\boldsymbol{k})$ represent real functions of $\boldsymbol{k}$, $\alpha^0 = \tau_0 \otimes \sigma_0$ and $\alpha^i = (\tau_x \otimes \sigma_0, \tau_y \otimes \sigma_z, \tau_z \otimes \sigma_z)$. Moreover, $\alpha^i$ satisfies the anticommutation relation, $\{\alpha^i, \alpha^j\} = 2\delta_{i,j}$, which guarantees a two-fold degeneracy. It permits the possible flavor Weyl points occurring at generic momenta when $d_i(\boldsymbol{k}) = 0$ for all $i$. Particularly, the two Weyl cones, with the basis $\{|A,\uparrow\rangle, |B,\uparrow\rangle\}$ and $\{|B,\downarrow\rangle, |A,\downarrow\rangle\}$, are degenerate due to $u_x^{1/2}$ symmetry.

While equation (2) does not have $SU(2)$ rotation symmetry of the Neel vector, there is a hidden $SU(2)$ symmetry protecting the flavor WSM phase. To elucidate this, we show the existence of a symmetry group with the group elements written as $exp(-i\theta \boldsymbol{n}\cdot\boldsymbol{\rho})$, where $\theta$, $\boldsymbol{n}$ and $\boldsymbol{\rho}$ represent the rotation angle, rotation axis, and three-vector of the generators of $SU(2)$ group, respectively. The generators can be constructed proportional to the representations of $u_x^{1/2}$, $u_y^{1/2}$, and $\{U_z(\pi)||E|0\}$, i.e., $\boldsymbol{\rho} =$ $\left(\frac{1}{2}\tau_x \otimes \sigma_x, \frac{1}{2}\tau_x \otimes \sigma_y, \frac{1}{2}\tau_0 \otimes \sigma_z\right)$. Since $\rho_i$ satisfies Pauli algebra, the group $\{exp(-i\theta \mathbf{n} \cdot \boldsymbol{\rho})\}$ is isomorphic to an $SU(2)$ symmetry group. Such a $SU(2)$ symmetry transforms the two Weyl cones $\{|A,\uparrow\rangle, |B,\uparrow\rangle\}$ and $\{|B,\downarrow\rangle, |A,\downarrow\rangle\}$, to their arbitrary linear combinations, as shown in Figure 2B. Since the Hamiltonian equation (2) can be diagonalized into spin-up and spin-down blocks, such $SU(2)$ group mixes spin-up and spin-down Weyl fermions to their linear combinations. This implies that the spin-up and spin-down Weyl fermions must have the same chirality, rendering the role of the $SU(2)$ symmetry as isospin symmetry connecting two Weyl fields with the same chirality.

The $SU(2)$ symmetry transforms spin and sublattice degrees of freedom simultaneously, leading to two degenerate states with distinct spatial wave functions differentiated by a sublattice transformation. Consequently, the surface spectra could be either nondegenerate or degenerate, depending on whether the surfaces break the sublattice transformation symmetry $u_x^{1/2}$ (Figure 1D). Thus, the $SU(2)$ symmetry presented in our spinful model is drastically different from the trivial $SU(2)$ spin rotation in nonmagnetic materials without SOC, which also supports charge-2 Weyl fermions by directly multiplying the spin index onto a spinless Weyl model.[31] Moreover, the flavor WSM protected by hidden $SU(2)$ symmetry under the regime of spin group has two distinct features compared with DSMs. i) Nonzero even Chern number (e.g., $C = \pm 2n$). Recall that $PT$ symmetry in DSMs guarantees two degenerate Weyl cones with opposite chirality, leading to a zero Chern number. In contrast, $\{exp(-i\theta \mathbf{n} \cdot \boldsymbol{\rho})\}$ ensures that the degenerate states have the same Berry curvature. ii) The flavor Weyl point exists in such collinear AFM systems without the protection of any additional symmetry except $u_x^{1/2}$. A perturbation to $H(\mathbf{k})$—that does not break $u_x^{1/2}$ or $U_z(\theta)$—typically shifts the position of the flavor Weyl point without opening a gap, resembling the case of Weyl semimetals.

**Material realization: CoNb₃S₆**

To realize flavor WSM in realistic materials, we first summarize the required conditions as following design principles: (i) collinear AFM order, (ii) broken $P$ and $PT$ symmetry, and (iii) presence of $\{T|\boldsymbol{\tau}_{1/2}\}$ symmetry. We note that (i) and (iii) ensure the presence of spin group symmetry $u_{x/y}^{1/2}$ and $U_z(\theta)$ without SOC. Based on these principles, we propose that the chiral transition-metal chalcogenide CoNb₃S₆ is a representative flavor WSM that hosts flavor Weyl points around the Fermi level. Based on these principles, we survey magnetic database in Bilbao server and find 443 crystalline materials (out of 1605 entries) that could host $SU(2)$ isospin symmetry; 62 of them could have flavor Weyl points when turning off SOC (see Supplementary Note S5 for the full list and the calculation of another material candidate GdCuSn). Among them, transition-metal chalcogenide CoNb₃S₆ is a representative flavor WSM that hosts flavor Weyl points around the Fermi level. Figure 3A shows that CoNb₃S₆ crystallizes in the chiral space group $P6_322$.[32] It has an AFM order with magnetic moments directed along a crystal axis within the a-b plane below the Neel temperature, $T_N$, of ~26 K.[33] This structure corresponds to the type-IV magnetic space group $P_B2_12_12$ (No. 18.22). It has one two-fold rotation along the *x*-axis $\{U_x(\pi)||C_x(\pi)|0\}$, two screw rotations $\{U_y(\pi)||C_y(\pi)|\boldsymbol{\tau}_{(b+c)/2}\}$ and $\{U_z(\pi)||C_z(\pi)|\boldsymbol{\tau}_{(b+c)/2}\}$, and nonsymmorphic time-reversal $\{T|\boldsymbol{\tau}_{(a+b)/2}\}$, as per the notation of spin space groups. Some symmetry operations beyond the conventional magnetic space group, including $\{E||C_x(\pi)|0\}$ and $\{TU_n(\pi)||E||0\}$ ($n \parallel cos\varphi\hat{y} + sin\varphi\hat{z}, \varphi \in (0,\pi]$), are permitted without SOC, forming the spin space group $P_B{}^12_1{}^12_1{}^12^{\infty m}1$ (see Supplementary Note S3).

In contrast to the previous calculations using nonmagnetic or alternative AFM configurations,[34,35] we adopt the experimental magnetic configuration observed by neutron scattering.[33] The band structure calculation (Figure 3C) shows that $CoNb_3S_6$ is a metal with multiple hole pockets near the $\Gamma$ point, consistent with the experiments showing holes as major system carriers (see Supplementary Note S4 for details about DFT calculation).[34] The symmetry properties guarantee the following topological features that appear in the band structure. First, the spin space group does not have $P$. However, it has $u_z^{1/2} = \{U_z(\pi)||E|\tau_{(a+b)/2}\}$ and $\{U_x(\theta)||E|0\}$, ensuring doubly degenerate bands for flavor WSM. Second, the two-fold spatial rotations decoupled to spin rotations—$\{E||C_x(\pi)|0\}$, $\{E||C_y(\pi)|\tau_{(a+c)/2}\}$ and $\{E||C_z(\pi)|\tau_{(a+c)/2}\}$—commute with $u_z^{1/2}$ along the $\Gamma - X$, $\Gamma - Y$, and $\Gamma - Z$ lines, ensuring that the two degenerate energy bands have identical rotation eigenvalues on the high-symmetry lines. Therefore, the three two-fold rotation operations can provide additional protection for flavor Weyl points. We note that although $CoNb_3S_6$ belongs to chiral space group, implying that all point-like degeneracies are chiral fermions, the occurrence of flavor Weyl nodes does not require a chiral space group in general.[36]

Remarkably, there are multiple flavor Weyl points around the Fermi level and four flavor Weyl points at ~0.7 eV above the Fermi level. The latter points are located along $\Gamma - X$ and $\Gamma - Y$ lines. We found that the crossing bands along these high-symmetry lines have opposite eigenvalues of $\{E||C_x(\pi)|0\}$ or $\{E||C_y(\pi)|\tau_{(a+c)/2}\}$, indicating that the flavor Weyl points are protected by $C_2$ rotation. The Berry curvature calculation (see

Figure 3D) shows that the two Weyl points along $(-X) - \Gamma - X$ act as the source of Berry curvature, and the other two act as the drain, manifesting their chiral nature. Further calculation of the Wilson loop showed that the Chern number over a spherical surface around a Weyl point along $\Gamma - X(Y)$ was -2 (+2) (see Supplementary Figure S1). Therefore, we name the flavor Weyl points along $\Gamma - X(Y)$ as $N_1, N_2$ ($P_1, P_2$). We obtained the Dirac-like $k \cdot p$ Hamiltonian in the following by applying the symmetry operations, $u_z^{1/2}$, $U_x(\theta)$ and $\{T||C_z(\pi)|\tau_{(b+c)/2}\}$, to the low-energy Hamiltonian near $N_1$:

$$H(\boldsymbol{k}) = (a_0 + a_1 k_x)\tau_0 \otimes \sigma_0 + (a_2 k_y)\tau_x \otimes \sigma_0 + (a_3 k_z)\tau_y \otimes \sigma_x + (a_4 k_x)\tau_z \otimes \sigma_x. \quad (3)$$

The results of our DFT calculation can be used to obtain the parameters of equation (3), giving rise to an anisotropic Dirac cone. By implementing the spin rotation $e^{-i1/2(\pi/2)\sigma_y}$ to equation (3) (transforming $\sigma_x$ terms into $\sigma_z$ terms), the Hamiltonian is block-diagonalized into two Weyl Hamiltonians of the same chirality.

The topological charges of the flavor WSM imply the existence of Fermi arc surface states connecting two flavor Weyl points with opposite chirality. However, flavor Weyl points with opposite chirality are not connected by any symmetry owing to the lack of inversion symmetry, roto-inversion symmetry and their combinations with $T$ in the system. Therefore, we found an energy difference of 87 meV between $P_1$ and $N_1$. Moreover, flavor Weyl points, $P_1$ and $P_2$ ($N_1$ and $N_2$), are connected by a two-fold spatial rotation. Hence, they are located at the same energy. There are two disconnected electron Fermi pockets, separately enclosing $P_1$ and $P_2$, and two disconnected hole pockets, separately enclosing $N_1$ and $N_2$, for the (001) surface when Fermi energy exists between the two. Every electron pocket is connected to a hole pocket by a branch of Fermi arc surface states due to the enclosure of the different topological charges in electron and hole pockets, forming a network across the Brillouin zone (see Figure 4A). Interestingly, the surface states are also doubly degenerate because $\{U_z(\pi)||E|\tau_{(a+b)/2}\}$ is preserved on this surface (and so as the hidden $SU(2)$ symmetry), in sharp contrast to the conventional topological insulators or DSMs where the surface bands are spin-polarized and nondegenerate. The degenerate Fermi arc surface states are split into two branches for the (100) surface with broken symmetry of $\{U_z(\pi)||E|\tau_{(a+b)/2}\}$, as shown in Figure 4B. The various Fermi arc surface states are robust against perturbations, maintaining the collinear A-type AFM order in the absence of SOC. On the contrary, topological protection for the surface states on the conventional DSM does not exist.[37]

The Chern number of a 2D slice in the Brillouin zone changed in the multiples of 2 because the flavor Weyl points have chiralities of $\pm 2$. Figure 4C shows that the Chern number of the slice perpendicular to the x-axis changes as a function of $k_y$. The Chern number calculated on slice near Γ—between flavor Weyl points with opposite chirality—is $\pm 2$. The result is consistent with the Berry curvature calculation (Figure 3D), where Berry curvature flows from $N_1$ and $N_2$ to $P_1$ and $P_2$. Figure 4D shows the corresponding edge states with two branches of chiral surface states connecting the conduction and valence bands that are doubly degenerate at $SU(2)$-preserved edge and nondegenerate at $SU(2)$-broken edge, further validating the interplay between the Weyl points and the hidden $SU(2)$ symmetry. We note that the energies of the flavor Weyl points $P_{1,2}$ and $N_{1,2}$ (~0.7 eV) depend on the $U$ value (3 eV) of Co-3$d$ electrons we adopt for correlation effects. When $U$ is set to 1 eV, the energies of these points shift to ~0.5 eV, whose fermi-arc state might be observed by angle-resolved photoemission spectroscopy.

**Effects of spin-orbit coupling**

While SOC is a universal relativistic property existing in all materials, for most materials, even with strong SOC, e.g., 10-100 meV, its influence on the electronic structure is still limited compared with those caused by exchange splitting and crystal field, etc. Therefore, we can take the SOC-free Hamiltonian, which is described by spin group symmetry, as a good starting point to understand magnetic materials with SOC by treating SOC as a perturbation that breaks certain spin group symmetries. Specifically, since the flavor Weyl points are charge-2 monopoles of Berry curvature, the sub-Hilbert space on a spherical surface encircling a flavor Weyl point should be a Chern insulator with Chern number $\pm 2$, which cannot be changed under any sort of symmetry-breaking perturbation, unless a gap-closing occurs in this sub-Heilberg space. Therefore, when SOC is included, although doubly degenerate bands split due to the broken $SU(2)$ isospin symmetry, the flavor Weyl point undergoes a phase transition to a twin-pair of conventional Weyl points with the same chirality rather than being gapped immediately.

We next study the modification of band dispersions in $CoNb_3S_6$, which depends on the specific bands and wavevectors, by turning on SOC. Figure 5A shows that the energy bands contributing to flavor Weyl points $P_{1,2}$ and $N_{1,2}$ have moderate spin splitting about 20 meV, while Figure 5D shows that most energy bands near the Fermi level have relatively small spin-splitting (<10 meV) in the presence of SOC. Thus, the SOC effects of the flavor Weyl points have two different manifestations, i.e., a twin-pair of Weyl points or fully gapped. For $P_{1,2}$ and $N_{1,2}$ (Figure 5A), because of the small energy gap around the loop in the Brillion zone passing these flavor Weyl points in the absence of SOC, SOC is large enough to gap these Weyl points. However, despite the gapped phase and spin-split surface states, the features of Fermi arc still resemble those without SOC, as shown in Figure 5B,C. The difference is that the Fermi-arc surface states are now trivial rather than nontrivial, connecting a electron (hole) pocket with a electron (hole) pocket. Recall the successful measurement of the Fermi-arc states in DSMs, such spin-group induced feature could also be visible for experiments. For flavor Weyl points near the Fermi level (Figure 5D), small spin splitting cause some flavor Weyl points to split into twin-pair Weyl points rather than being gapped, as shown in Figure 5E,F. The spin splitting at the Weyl points is only ~3 meV, which is a small perturbation to the flavor Weyl points protected by spin group.[38-41]

Overall, even if SOC effect is generally not negligible in $CoNb_3S_6$, the flavor WSM phases can still be considered as a starting point to understand its topological nature that cannot be fully described by magnetic space group. Interestingly, the SOC-free approximation of the flavor symmetry studied here also makes a nice analogy to the flavor symmetry in particle physics, which is also an approximate symmetry. Recall that isospin symmetry is good enough in prediction of the possibility and rates of nuclear reaction when the masses of the two particles, e.g., proton (938.27 MeV) and neutron (939.57 MeV), are similar, spin group symmetry protects degeneracies, topological charges, and surface states of certain topological materials when SOC is weak.[42]

**DISCUSSION**

We discuss the possible experimental phenomena associated with flavor Weyl fermions. First, flavor WSM manifests unique and robust surface states enforced by topological charges. On $SU(2)$ preserved surfaces, the surface states would be Weyl-like, while on $SU(2)$ breaking surfaces, the surface states would be Dirac-like. In contrast, robust surface states of DSMs are rare except for specific nonsymmorphic symmetries to protect the surface states.[37,43] Owing to the protection of topological charge, even if sizable SOC exists, flavor Weyl points could be split into a twin-pair of conventional Weyl points with the same chirality rather than being gapped immediately. These characteristics are potentially observable by angle-resolved photoemission spectrocopy.

More importantly, the robust fermi-arc surface states of flavor WSM potentially lead to unexplored emergent transport and optical properties. For example, the flavor Weyl points of opposite chirality in $CoNb_3S_6$ do not lie at the same energy, possibly leading to a large and quantized response to circularly polarized light.[44] Furthermore, the net anomalous Hall conductivity and spin Hall conductivity in $CoNb_3S_6$ should be zero owing to the presence of $\{T||E|\boldsymbol{\tau}_{1/2}\}$ symmetry and $\{T||U_{\boldsymbol{n}}(\pi)|0\}$ symmetry. However, breaking $\{T||E|\boldsymbol{\tau}_{1/2}\}$ symmetry and $\{T||U_{\boldsymbol{n}}(\pi)|0\}$ symmetry through the small SOC effect and small tilting of magnetic moments may lead to a large anomalous Hall conductance because of the uncompensated Berry curvature and multiple Fermi arcs emerging from the charge-2 flavor Weyl points. It has been observed in $CoNb_3S_6$, accompanied by small out-of-plane components of the magnetic moments.[34]

Poincare symmetry is generally broken in solid-state lattices, while certain crystalline symmetries such as nonsymmorphic symmetry are absent in high-energy physics. Because of these differences, there are various types of quasiparticle excitation in condensed matter physics that do not have counterparts in high-energy physics, including three-, six- or eight-fold degenerate points,[45,46] line-like,[47-49] chain-like,[50,51] and plane-like band crossings,[52] etc. Besides, there are also emergent quasiparticles composed of two Weyl points of opposite chirality, like Dirac fermions, but with different velocities, indicating that energy bands around the four-fold degenerate points are generally nondegenerate.[53,54] We note that such quasiparticles sometimes are also attributed to a type of DSM with a looser definition, which allows band splitting around Dirac points.[5] In addition, previous literature also reported four-component Weyl fermions with nonzero Chern number $\pm 2$ or $\pm 4$, in both electron[5,22,23,55-57] and phonon systems.[58,59] We note that the main difference between these quasiparticles and the flavor Weyl fermions in the presented work is twofold. First, these quasiparticles are stabilized by the little groups with high-order rotation operations or the little groups with nonsymmorphic symmetry operations. Therefore, these elementary excitations can only appear at specific high-symmetry momenta of the Brillouin zone. However, the flavor Weyl points can appear at generic momenta. This property implies the emergence of dense flavor Weyl points within a small energy range, possibly leading to stronger topological effects. Second, the previously studied four-fold degenerate points with nonzero Chern number inevitably have nondegenerate energy bands away from the high-symmetry points. Therefore, they do not strictly fulfill the massless four-component equation in quantum field theory. However, our flavor WSM model was derived from the quantum field theory perspective with doubly degenerate dispersions around the Dirac-like points, stabilized by the hidden $SU(2)$ isospin symmetry.

**REFERENCES**

1. Novoselov, K.S., Geim, A.K., Morozov, S.V., et al. (2005). Two-dimensional gas of massless Dirac fermions in graphene. Nature **438**, 197-200.

2. Fu, L., Kane, C.L., and Mele, E.J. (2007). Topological insulators in three dimensions. Phys. Rev. Lett. **98**, 106803.

3. Moore, J.E., and Balents, L. (2007). Topological invariants of time-reversal-invariant band structures. Phys. Rev. B *75*, 121306.

4. Wang, P., Ge, J., Li, J., et al. (2021). Intrinsic magnetic topological insulators. The Innovation 2(2), 100098.

5. Young, S.M., Zaheer, S., Teo, J.C.Y., et al. (2012). Dirac semimetal in three dimensions. Phys. Rev. Lett. *108*, 140405.

6. Yang, B.-J., and Nagaosa, N. (2014). Classification of stable three-dimensional Dirac semimetals with nontrivial topology. Nat. Commun. *5*, 4898.

7. Young, S.M., and Kane, C.L. (2015). Dirac semimetals in two dimensions. Phys. Rev. Lett. *115*, 126803.

8. Wan, X., Turner, A.M., Vishwanath, A., and Savrasov, S.Y. (2011). Topological semimetal and fermi-arc surface states in the electronic structure of pyrochlore iridates. Phys. Rev. B *83*, 205101.

9. Fang, C., Gilbert, M.J., Dai, X., and Bernevig, B.A. (2012). Multi-Weyl topological semimetals stabilized by point group symmetry. Phys. Rev. Lett. *108*, 266802.

10. Armitage, N.P., Mele, E.J., and Vishwanath, A. (2018). Weyl and Dirac semimetals in three-dimensional solids. Rev. Mod. Phys. *90*, 015001.

11. Wang, Z., Sun, Y., Chen, X.-Q., et al. (2012). Dirac semimetal and topological phase transitions in $A_3$Bi ($A$= Na, K, Rb). Phys. Rev. B *85*, 195320.

12. Wang, Z., Weng, H., Wu, Q., et al. (2013). Three-dimensional Dirac semimetal and quantum transport in $Cd_3As_2$. Phys. Rev. B *88*, 125427.

13. Liu, Z.K., Jiang, J., Zhou, B., et al. (2014). A stable three-dimensional topological Dirac semimetal $Cd_3As_2$. Nat. Mater. *13*, 677-681.

14. Liu, Z.K., Zhou, B., Zhang, Y., et al. (2014). Discovery of a three-dimensional topological Dirac semimetal, Na$_3$Bi. Science *343*, 864-867.

15. Tang, P., Zhou, Q., Xu, G., and Zhang, S.-C. (2016). Dirac fermions in an antiferromagnetic semimetal. Nat. Phys. *12*, 1100-1104.

16. Wang, J. (2017). Antiferromagnetic Dirac semimetals in two dimensions. Phys. Rev. B *95*, 115138.

17. Hua, G., Nie, S., Song, Z., et al. (2018). Dirac semimetal in type-IV magnetic space groups. Phys. Rev. B *98*, 201116.

18. Watanabe, H., Po, H.C., and Vishwanath, A. (2018). Structure and topology of band structures in the 1651 magnetic space groups. Sci. Adv. *4*, eaat8685.

19. Xu, Y., Elcoro, L., Song, Z.-D., et al. (2020). High-throughput calculations of magnetic topological materials. Nature *586*, 702-707.

20. Elcoro, L., Wieder, B.J., Song, Z., et al. (2021). Magnetic topological quantum chemistry. Nat. Commun. *12*, 1-10.

21. Peng, B., Jiang, Y., Fang, Z., et al. (2022). Topological classification and diagnosis in magnetically ordered electronic materials. Phys. Rev. B *105*, 235138.

22. Yu, Z.-M., Zhang, Z., Liu, G.-B., et al. (2022). Encyclopedia of emergent particles in three-dimensional crystals. Sci. Bull. *67*, 375-380.

23. Tang, F., and Wan, X. (2021). Exhaustive construction of effective models in 1651 magnetic space groups. Phys. Rev. B *104*, 085137.

24. Brinkman, W.F., and Elliott, R.J. (1966). Theory of spin-space groups. Proc. R. Soc. A *294*, 343-358.

25. Litvin, D.B., and Opechowski, W. (1974). Spin groups. Physica *76*, 538-554.

26. Liu, P., Li, J., Han, J., et al. (2022). Spin-group symmetry in magnetic materials with negligible spin-orbit coupling. Phys. Rev. X *12*, 021016.

27. Yang, J., Liu, Z.-X., and Fang, C. (2021). Symmetry invariants of spin space groups in magnetic materials. arXiv:2105.12738.

28. Corticelli, A., Moessner, R., and McClarty, P.A. (2022). Spin-space groups and magnon band topology. Phys. Rev. B *105*, 064430.

29. Coleman, S., and Mandula, J. (1967). All possible symmetries of the s matrix. Phys. Rev. *159*, 1251.

30. Heisenberg, W. (1932). Über den bau der atomkerne. I. Z. Phys. *77*, 1-11.

31. Chang, G., Xu, S.-Y., Sanchez Daniel, S., et al. (2016). A strongly robust type II Weyl fermion semimetal state in $Ta_3S_2$. Sci. Adv. *2*, e1600295.

32. Anzenhofer, K., Van Den Berg, J.M., Cossee, P., and Helle, J.N. (1970). The crystal structure and magnetic susceptibilities of $MnNb_3S_6$, $FeNb_3S_6$, $CoNb_3S_6$ and $NiNb_3S_6$. J. Phys. Chem. Solids *31*, 1057-1067.

33. Parkin, S.S.P., Marseglia, E.A., and Brown, P.J. (1983). Magnetic structure of $Co_{1/3}NbS_2$ and $Co_{1/3}TaS_2$. J. Phys. C: Solid State Phys. *16*, 2765.

34. Ghimire, N.J., Botana, A.S., Jiang, J.S., et al. (2018). Large anomalous hall effect in the chiral-lattice antiferromagnet $CoNb_3S_6$. Nat. Commun. *9*, 3280.

35. Šmejkal, L., González-Hernández, R., Jungwirth, T., and Sinova, J. (2020). Crystal time-reversal symmetry breaking and spontaneous hall effect in collinear antiferromagnets. Sci. Adv. *6*, eaaz8809.

36. Chang, G., Wieder, B.J., Schindler, F., et al. (2018). Topological quantum properties of chiral crystals. Nat. Mater. *17*, 978-985.

37. Kargarian, M., Randeria, M., and Lu, Y.-M. (2016). Are the surface fermi arcs in Dirac semimetals topologically protected? Proc. Natl. Acad. Sci. U.S.A. *113*, 8648-8652.

38. Yi, H., Wang, Z., Chen, C., et al. (2014). Evidence of topological surface state in three-dimensional Dirac semimetal $Cd_3As_2$. Sci. Rep. *4*, 6106.

39. Xu, S.-Y., Liu, C., Kushwaha, S.K., et al. (2015). Observation of fermi arc surface states in a topological metal. Science *347*, 294-298.

40. Moll, P.J.W., Nair, N.L., Helm, T., et al. (2016). Transport evidence for fermi-arc-mediated chirality transfer in the Dirac semimetal $Cd_3As_2$. Nature *535*, 266-270.

41. Wu, Y., Jo, N.H., Wang, L.-L., et al. (2019). Fragility of fermi arcs in Dirac semimetals. Phys. Rev. B *99*, 161113.

42. Zee, A. (2016). Group theory in a nutshell for physicists (Princeton University Press).

43. Fang, C., Lu, L., Liu, J., and Fu, L. (2016). Topological semimetals with helicoid surface states. Nat. Phys. *12*, 936-941.

44. de Juan, F., Grushin, A.G., Morimoto, T., and Moore, J.E. (2017). Quantized circular photogalvanic effect in Weyl semimetals. Nat. Commun. *8*, 15995.

45. Bradlyn, B., Cano, J., Wang, Z., et al. (2016). Beyond Dirac and Weyl fermions: Unconventional quasiparticles in conventional crystals. Science *353*, aaf5037.

46. Wieder, B.J., Kim, Y., Rappe, A.M., and Kane, C.L. (2016). Double Dirac semimetals in three dimensions. Phys. Rev. Lett. *116*, 186402.

47. Burkov, A.A., Hook, M.D., and Balents, L. (2011). Topological nodal semimetals. Phys. Rev. B *84*, 235126.

48. Fang, C., Chen, Y., Kee, H.-Y., and Fu, L. (2015). Topological nodal line semimetals with and without spin-orbital coupling. Phys. Rev. B *92*, 081201.

49. Kim, Y., Wieder, B.J., Kane, C.L., and Rappe, A.M. (2015). Dirac line nodes in inversion-symmetric crystals. Phys. Rev. Lett. *115*, 036806.

50. Bzdušek, T., Wu, Q., Rüegg, A., et al. (2016). Nodal-chain metals. Nature *538*, 75-78.

51. Wu, Q., Soluyanov, A.A., and Bzdušek, T. (2019). Non-abelian band topology in noninteracting metals. Science *365*, 1273-1277.

52. Wu, W., Liu, Y., Li, S., et al. (2018). Nodal surface semimetals: Theory and material realization. Phys. Rev. B *97*, 115125.

53. Gao, H., Kim, Y., Venderbos, J.W.F., et al. (2018). Dirac-Weyl semimetal: Coexistence of Dirac and Weyl fermions in polar hexagonal *ABC* crystals. Phys. Rev. Lett. *121*, 106404.

54. Wieder, B.J., Wang, Z., Cano, J., et al. (2020). Strong and fragile topological Dirac semimetals with higher-order fermi arcs. Nat. Commun. *11*, 627.

55. Xu, Y., and Duan, L.M. (2016). Type-II Weyl points in three-dimensional cold-atom optical lattices. Phys. Rev. A *94*, 053619.

56. Chang, G., Xu, S.-Y., Wieder, B.J., et al. (2017). Unconventional chiral fermions and large topological fermi arcs in RhSi. Phys. Rev. Lett. *119*, 206401.

57. Tang, P., Zhou, Q., and Zhang, S.-C. (2017). Multiple types of topological fermions in transition metal silicides. Phys. Rev. Lett. *119*, 206402.

58. Zhang, T., Song, Z., Alexandradinata, A., et al. (2018). Double-Weyl phonons in transition-metal monosilicides. Phys. Rev. Lett. *120*, 016401.

59.     Miao, H., Zhang, T.T., Wang, L., et al. (2018). Observation of double Weyl phonons in parity-breaking FeSi. Phys. Rev. Lett. *121*, 035302.

**ACKNOWLEDGMENTS**

We thank Profs. Xiangang Wan, Huaxing Zhu and Sixue Qin for helpful discussions. This work was supported by National Key R&D Program of China under Grant No. 2020YFA0308900, the National Natural Science Foundation of China under Grant No. 11874195, Guangdong Innovative and Entrepreneurial Research Team Program under Grant No. 2017ZT07C062, Guangdong Provincial Key Laboratory for Computational Science and Material Design under Grant No. 2019B030301001, the Shenzhen Science and Technology Program (Grant No.KQTD20190929173815000) and Center for Computational Science and Engineering of Southern University of Science and Technology.

**AUTHOR CONTRIBUTIONS**

Q. L. supervised the research with J. H.. P. L. and A. Z. performed the calculations. All the authors contributed inputs to write the paper.

**DECLARATION OF INTERESTS**

The authors declare no competing interests.

**SUPPLEMENTAL INFORMATION**

Supplemental information can be found online at…

**LEAD CONTACT WEBSITE**

Qihang Liu: https://liuqh.phy.sustech.edu.cn/

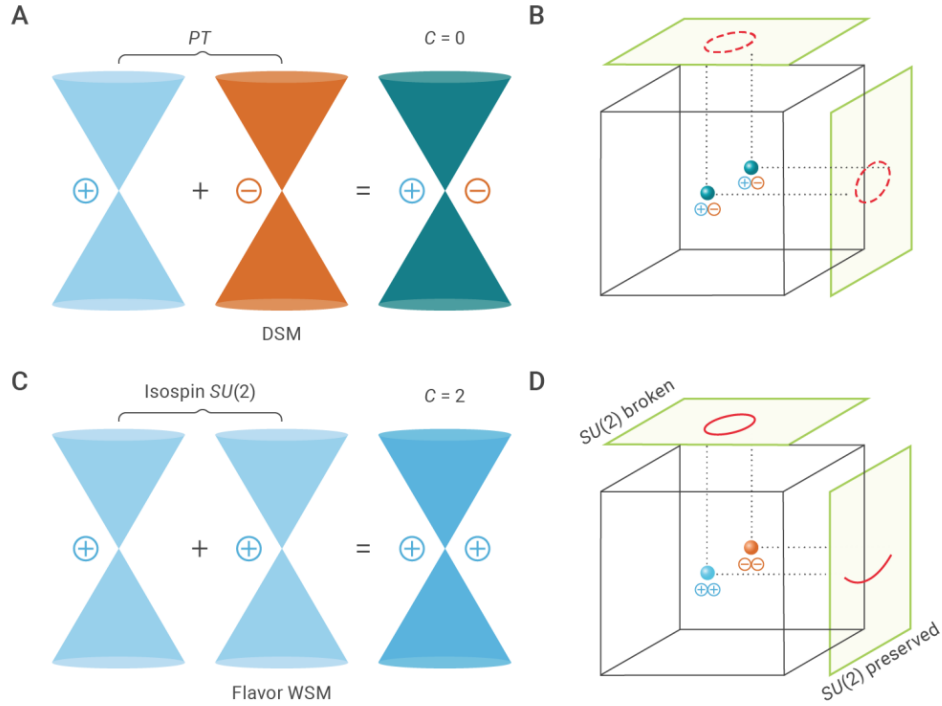

**Figure 1. Schematic of the Dirac semimetal (DSM) and flavor Weyl semimetal (flavor WSM).** (A) A Dirac point can be viewed as the superposition of two Weyl points with opposite chirality in a DSM. Such superposition is generally obtained by the space-time $PT$ symmetry. (B) The surface states of the DSM are adiabatically connected to topologically trivial surface states. The green points denote the Dirac points. (C) A flavor WSM hosts 4-fold degenerate points composed of two Weyl points with identical chirality, protected by a hidden $SU(2)$ symmetry group (analogous to the isospin symmetry in particle physics). (D) The surface states of flavor WSM are robust owing to the protection of chiral charges. The surface states on the surfaces that preserve the $SU(2)$ symmetry are two-fold degenerate connecting two flavor Weyl points with opposite chirality. However, the surface states on the surfaces with a broken $SU(2)$ symmetry group split into two spin-polarized branches, resembling conventional topological insulators or semimetals.

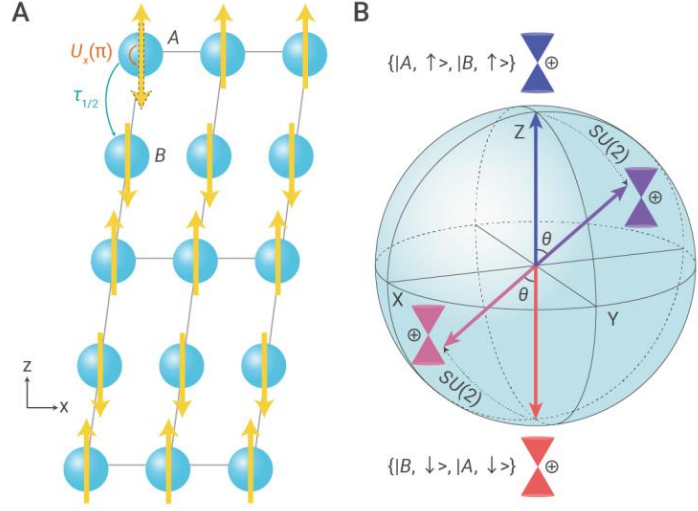

**Figure 2. Hidden $SU(2)$ symmetry in antiferromagnetic materials.** (A) The magnetic lattice with collinear antiferromagnetic order allows spin-group symmetry operations, $\{U_x(\pi)||E|\tau_{1/2}\}$ and $\{U_z(\theta)||E|0\}$ without spin-orbit coupling, leading to two degenerate Weyl cones with the basis $\{|A,\uparrow\rangle, |B,\uparrow\rangle\}$ and $\{|B,\downarrow\rangle, |A,\downarrow\rangle\}$ and a $SU(2)$ symmetry group $\{exp(-i\theta \boldsymbol{n}\cdot\boldsymbol{\rho})\}$ (see the main text). (B) Bloch sphere of the $SU(2)$ symmetry group, transforming the basis of a Weyl cone $\{|A,\uparrow\rangle, |B,\uparrow\rangle\}$ (blue arrow) to any linear combinations (up to a phase factor) $\{\alpha|A,\uparrow\rangle + \beta|B,\downarrow\rangle, \alpha|B,\uparrow\rangle + \beta|A,\downarrow\rangle\}$, and transforming $\{|B,\downarrow\rangle, |A,\downarrow\rangle\}$ (red arrow) to an orthogonal one $\{-\beta^*|A,\uparrow\rangle + \alpha^*|B,\downarrow\rangle, -\beta^*|B,\uparrow\rangle + \alpha^*|A,\downarrow\rangle\}$. The basis transformation under the rotation axis (grey line) $\boldsymbol{n} = (cos(\omega), sin(\omega), 0)$ and rotation angle $\theta$ are also shown. The mixing coefficients are $\alpha = cos[\theta/2]$ and $\beta = -isin[\theta/2]e^{-i\omega}$.

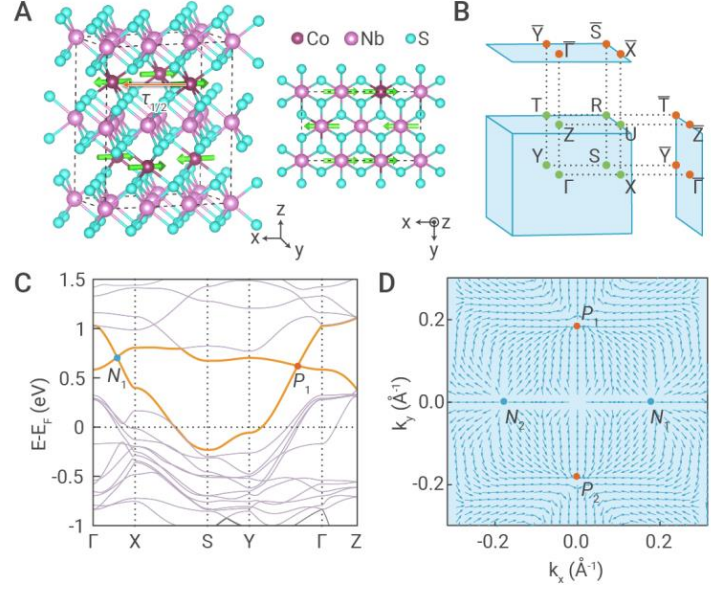

**Figure 3. Crystal and bulk electronic properties of CoNb$_3$S$_6$.** (A) The crystal structure of CoNb$_3$S$_6$. (B) The Brillouin zones of bulk (001) and (100) surfaces of CoNb$_3$S$_6$. (C) The band structure of CoNb$_3$S$_6$ without spin-orbit coupling. There are two flavor Weyl points at ~0.7 eV above the fermi level, $N_1$ and $P_1$, and another two flavor Weyl points, $N_2$ and $P_2$ (not shown), that are connected to $N_1$ and $P_1$ through two-fold rotation. (D) Distribution of in-plane components of the trace of Berry curvature tensor on $k_z = 0$ plane, where $N_1/N_2$ and $P_1/P_2$ denote the source and sink, respectively.

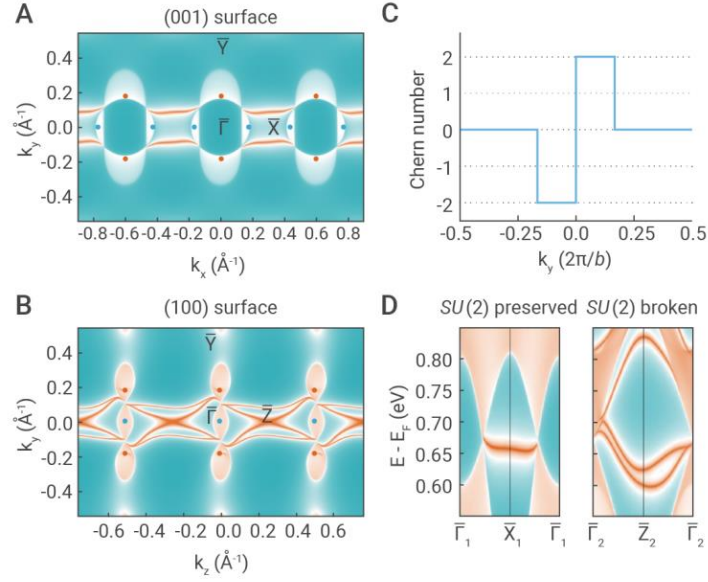

**Figure 4. Protected topological surface states of flavor WSM CoNb$_3$S$_6$.** (A), (B) Iso-energy surface states connect electron pockets and hole pockets, separately enclosing flavor Weyl points with opposite chirality on (001) and (100) surfaces of CoNb$_3$S$_6$. (C) The transition of Chern number defined on 2D slices in the Brillouin zone perpendicular to the y-axis as a function of momentum $k_y$. (D) Chiral edge states of the 2D slice ($k_y = 0.1$ ($2\pi/b$)) with Chern number of 2. Doubly degenerate edge bands are found in $SU(2)$-preserved edge, while spin-polarized nondegenerate edge bands are found in $SU(2)$-broken edge. The notations are defined as $\bar{\Gamma}_1 = \bar{\Gamma} + \bar{P}$, $\bar{X}_1 = \bar{X} + \bar{P}$, $\bar{\Gamma}_2 = \bar{\Gamma} + \bar{P}$, and $\bar{Z}_2 = \bar{Z} + \bar{P}$, where $\bar{P} = \frac{1}{5}(\bar{Y} - \bar{\Gamma})$.

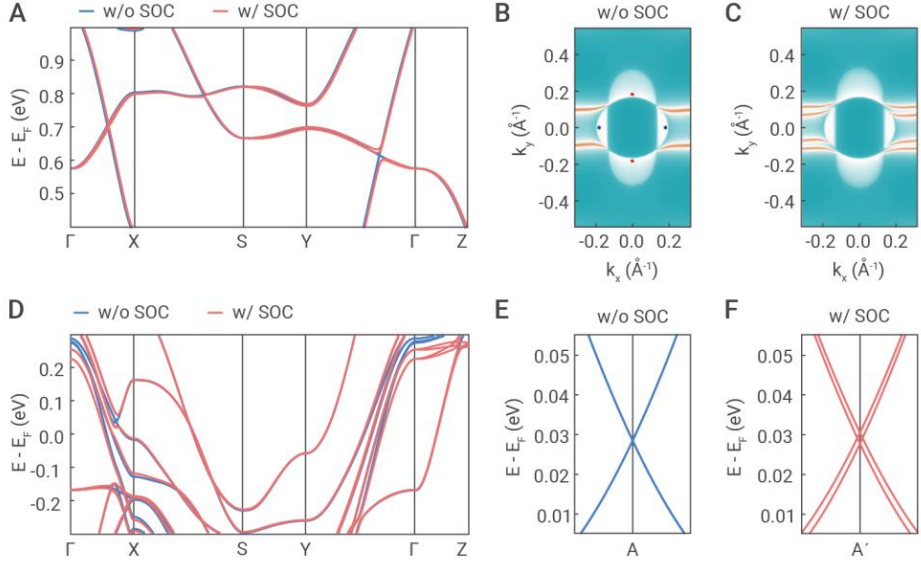

**Figure 5. Effects of spin-orbit coupling on energy bands.** (A) Band structure around the energies of $N_1$ and $P_1$. (B), (C) Iso-energy topological surface states without SOC (B) and with SOC (C). The energy is set between those of $N_1$ and $P_1$. (D) Band structure around the Fermi level. (E), (F) Zoom-in bands of a flavor Weyl point without SOC (E) and with SOC (F). The coordinate of points labeled are $A = (0.5, 0.2572, 0), A' = (0.5, 0.2575, 0), \delta = (0.05, 0, 0)$.